\documentclass[pre,twocolumn,superscriptaddress,amsmath,amssymb,showpacs]{revtex4}
\usepackage{fancyhdr}
\usepackage{graphicx}
\usepackage{amsmath}
\usepackage{bm}
\usepackage{subfigure}
\usepackage{lastpage}
\newcommand{\bra}[1]{\langle #1|}
\newcommand{\ket}[1]{|#1\rangle}

\begin{document}

\title{The independent scattering approximation in the saturated regime}

\author{Tom \surname{Savels}}
\email{t.savels@amolf.nl}\affiliation{\small{FOM Institute for
Atomic and Molecular Physics, Kruislaan 407, 1098 SJ, Amsterdam,
The Netherlands}} \affiliation{\small{Complex Photonic Systems,
Dept. Science and Technology\\\small University of Twente, PO Box
217, 7500 AE Enschede, The Netherlands.}}
\author{Allard P. \surname{Mosk}}
\affiliation{\small{Complex Photonic Systems, Dept. Science and
Technology\\\small University of Twente, PO Box 217, 7500 AE
Enschede, The Netherlands.}}
\author{Ad \surname{Lagendijk}}
\affiliation{\small{FOM Institute for Atomic and Molecular
Physics, Kruislaan 407, 1098 SJ, Amsterdam, The Netherlands}}
\affiliation{\small{Complex Photonic Systems, Dept. Science and
Technology\\\small University of Twente, PO Box 217, 7500 AE
Enschede, The Netherlands.}}

\begin{abstract}
We show that a saturable single-frequency elastic T-matrix
approach to scattering of light by atoms agrees remarkably well
with a master equation description in the regime of unsaturated
atoms, or for large separation between the atoms. If the atoms are
in each other's near-field and the saturation of the atoms is high
enough, the two approaches yield different results.
\end{abstract}

\maketitle \lfoot{} \fancyhead[RO]{} \cfoot{} \pagestyle{fancy}

\section{Introduction}
Applications of multiple scattering of light by point-scatterers
are often based on the assumption that the scatterers under
consideration scatter light independently from one another. All
the scattering properties of each individual scatterer can then be
expressed by a single mathematical object called the scatterer's
T-matrix \cite{B v Tiggelen, Th Nieuwenhuizen}. This
single-scatterer T-matrix is used as a building block in
multiple-scattering theories \cite{C Bohren} which can accurately
describe many interesting phenomena \cite{M Havey, M Mark}.
However, the use of T-matrices usually implies that one considers
only elastic scattering. Although the restriction of scattering
events to the pure elastic ones is justifiable in some cases, it
is not necessarily correct for, e.g., scattering of high-intensity
incident light fields \cite{T Chaneliere}.\\
\indent In this paper, we will consider scattering of light by two
atoms. The ``atoms" could be implemented as any type of
sub-wavelength quantum objects, for example: trapped atoms,
quantum dots \cite{D Gammon}, trapped ions \cite{D Leibfried} or dye molecules. If we model the atoms
as point-dipoles, we can use the T-matrix formalism to describe
scattering of incident light. Alternatively, we can model the
atoms as two-level systems, in which case multiple scattering of
light can be studied using a master equation \cite{C
Cohen-Tannoudji, M Scully, H Carmichael}. The system is then
characterized by a density matrix, representing the coherences and
populations of the atomic levels. Scattering of incident light then
determines the evolution of the density matrix, while taking into
account all orders of multiple scattering as well as inelastic
scattering of light.\\
\indent The goal of this paper is to compare the scattering
properties of two atoms as described by a simple T-matrix approach
and a (computationally much more involved) master equation
approach. We will show that a saturable single-frequency elastic
T-matrix approach agrees remarkably well with a master equation
description in the regime of unsaturated atoms or for
large separation between the atoms.\\
\indent The new aspect of this work is the comparison of two
approaches which seem completely different at first but are in
fact closely connected. We stress that the two approaches
themselves, reviewed for clarity in sections II and III of our
paper, have been extensively studied previously (see, e.g.,
\cite{B v Tiggelen, C Cohen-Tannoudji})
\\
\indent This paper is organized as follows. We start in Section II
by describing the system we are considering. A master equation
approach allows us to calculate how a near-resonant monochromatic
incident field is scattered. In Section III, we consider the same
system, and use a T-matrix approach to determine the scattering
properties of the system. In Section IV we compare both methods.
In Section V, finally, we discuss the applicability of our results
to other atomic systems.
\section{The density matrix approach to scattering of light}
We start by considering the system shown in Figure 1.
\begin{figure}
    \begin{center}\includegraphics[width=3in]{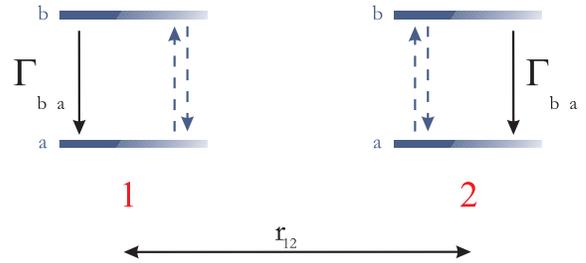}
    \caption{\label{fig,setup} The system under consideration: two identical
    atoms
    $1$ and $2$ are positioned in free space at a distance $r_{12}$ from each other.
    Both atoms have a two-level internal structure. The upper level has a lifetime
    $\Gamma_{ba}^{-1}$. Both atoms interact with an incident field; this interaction
    is depicted by the dashed arrows.}
    \end{center}
\end{figure}
Two identical atoms are positioned in free space, separated by a
distance $r_{12}=|{\bm r }_{12}|$. Both atoms are modelled as a
two-level system with upper level $b$ and lower level $a$,
separated by an energy difference $\hbar\omega_{ba}$. Decay
between both levels occurs according to the decay rate
$\Gamma_{ba}$. A transition dipole moment ${\bm d}_{ab}^{i}$,
$i\in\{1,2\}$ is associated with both $b\rightarrow a$
transitions. Both transition dipole moments are chosen to be equal
in magnitude and orientation:
\\
\begin{align}
{\bm d}_{ab}^{i}\equiv d_{ab}{\bm \mu},\qquad |{\bm \mu}|=1,\qquad
i\in\{1,2\}.\label{dipole moments}
\end{align}
\\
\indent An incident monochromatic field $ {\bm E}_{0} $ with wave
vector ${\bm k}_{0}$ interacts with both atoms. The frequency
$\omega_{0}=|{\bm k}_{0}|c$ of the incident field is tunable and
chosen near the $b\rightarrow a$ resonance ($c$ is the velocity of
light in free space). For simplicity, we have taken the wave
vector ${\bm k}_{0}$ of the incident field perpendicular to ${\bm
r}_{12}$. This choice of ${\bm k}_{0}$ ensures that both atoms
always feel the same phase of the incident field, which makes the
calculations somewhat
easier without affecting the generality of our results. \\
\indent We will now derive the master equation of the system; the
following derivation closely follows \cite{R Lehmberg}. The total
Hamiltonian $\hat{H}_{0}$ describing the energies of the systems,
the electromagnetic field and interactions is, in the standard
electric dipole and rotating wave approximation, given by
\\
\begin{align}
\hat{H}_{0}\equiv\hat{H}_{D}+\hat{H}_{F}+\hat{H}_{DL}+\hat{H}_{DF}\label{H_0}.
\end{align}
\\
\indent The Hamiltonian of both individual atoms is given by
\\
\begin{align}
\hat{H}_{D}\equiv\hbar\omega_{ba}(\hat{S}_{1}^{+}\hat{S}_{1}^{-}+\hat{S}_{2}^{+}\hat{S}_{2}^{-}),
\end{align}
\\
with $\hat{S}_{i}^{\pm}$ the dipole raising and lowering operators
of atom $i$. Both atoms are coupled to the three-dimensional
multimode electromagnetic field with Hamiltonian
\\
\begin{align}
\hat{H}_{F}\equiv\sum_{{\bm k}\lambda}\hbar\omega_{{\bm
k}\lambda}\Bigl(\hat{a}_{{\bm k}\lambda}^{\dagger}\hat{a}_{{\bm
k}\lambda}+\frac{1}{2}\Bigr),
\end{align}
\\
where the operators $\hat{a}_{{\bm k}\lambda}^{\dagger}$ and
$\hat{a}_{{\bm k}\lambda}$ respectively create and annihilate a
photon in de mode $({\bm k},\lambda)$. The modes of the
electromagnetic field are taken to be in a vacuum state. Both
atoms also interact through the Hamiltonian
\\
\begin{align}
\hat{H}_{DL}\equiv
\frac{1}{2}\hbar\Omega\Bigl((\hat{S}_{1}^{+}+\hat{S}_{2}^{+})e^{i\omega_{0}
t }+\text{H.c.}\Bigr),\label{H_DL}
\end{align}
\\
with an incident field ${\bm E}_{0}$ with frequency $\omega_{0}$,
where the Rabi frequency is given by $\Omega\equiv |{\bm
d}_{ab}\cdot {\bm E}_{0}|/\hbar$. Finally, the atoms interact with
the multimode vacuum field through the Hamiltonian
\\
\begin{align}
\hat{H}_{DF}\equiv\sum_{{\bm k}\lambda}\Bigl({\bm \mu}\cdot({\bm g
}_{{\bm k}\lambda}({\bm r}_{1})\hat{S}_{1}^{+}+{\bm g }_{{\bm
k}\lambda}({\bm r}_{2})\hat{S}_{2}^{+})+\text{H.c.}\Bigr),
\end{align}
\\
where the mode function ${\bm g}_{{\bm k\lambda}}$ is given by
\\
\begin{align}
{\bm g}_{{\bm k}\lambda}({\bm
r})\equiv\sqrt{\frac{\hbar\omega_{{\bm
k\lambda}}}{2\varepsilon_{0}L^3}}e^{i{\bm k}\cdot{\bm r}}{\bm
\varepsilon}_{{\bm k\lambda}},
\end{align}
\\
with $L^3$ the quantization volume, $\varepsilon_{0}$ the vacuum
permittivity and ${\bm \varepsilon}_{{\bm k\lambda}}$ the unit
polarization vector of the field mode $({\bm
k},\lambda)$.\\
\indent The dynamics of the total system can be expressed in terms
of the Master equation for the density matrix $\hat{\sigma}$. The
Master equation is, in the standard Born and Markov approximation,
written in the Lindblad form, given by \cite{R Lehmberg, G Lindblad}
\\
\begin{align}
\frac{d}{dt}\hat{\sigma}&=\hat{\mathcal{L}}\hat{\sigma}\nonumber\\
&=\hat{\mathcal{L}}_{nd}\hat{\sigma}+\hat{\mathcal{L}}_{d}\hat{\sigma}\label{Master,Lindblad}.
\end{align}
\\
The non-dissipative part of the Lindblad operator can be written
as
\\
\begin{align}
    \hat{\mathcal{L}}_{nd}\hat{\sigma}& \equiv -\frac{i}{\hbar}[\hat{H}_{C}+\hat{H}_{0},\hat{\sigma}],\label{sec,Master,Mastereq_NonDissipative}
\end{align}
\\
with coupling Hamiltonian
\\
\begin{align}
\hat{H}_{C}\equiv\hbar\delta^{(12)}\Bigl(\hat{S}_{1}^{+}\hat{S}_{2}^{-}+\hat{S}_{2}^{+}\hat{S}_{1}^{-}\Bigr),\label{nondissipative}
\end{align}
\\
while the dissipative part of the Master equation is given by
\\
\begin{align}
    \hat{\mathcal{L}}_{d}\hat{\sigma} \equiv&
    -\frac{\Gamma_{ba}}{2}\left(\hat{S}_{1}^{+}\hat{S}_{1}^{-}\hat{\sigma}-\hat{S}_{1}^{-}\hat{\sigma}\hat{S}_{1}^{+}+\hat{S}_{2}^{+}\hat{S}_{2}^{-}\hat{\sigma}-\hat{S}_{2}^{-}\hat{\sigma}\hat{S}_{2}^{+}\right)\nonumber\\
    &-\frac{\Gamma^{(12)}}{2}\left(\hat{S}_{1}^{+}\hat{S}_{2}^{-}\hat{\sigma}+\hat{\sigma}\hat{S}_{1}^{+}\hat{S}_{2}^{-}-2\hat{S}_{2}^{-}\hat{\sigma}\hat{S}_{1}^{+}\right)+\text{H.c.}\label{dissipative}
\end{align}
\\
It is clear from equations (\ref{nondissipative}) and
(\ref{dissipative}) that the vacuum induces coupling between both
atoms \cite{U Akram, D Cardimona, P Milonni, P Milonni 2}. The
parameters $\delta^{(12)}$ and $\Gamma^{(12)}$ are not associated
with individual systems, but with the total system as a whole. The
cross-damping $\Gamma^{(12)}$ represents incoherent coupling
between both atoms through spontaneous emission
\\
\begin{align}
\Gamma^{(12)}=\frac{2\pi}{\hbar^2}\sum_{{\bm k}\lambda}\Bigl({\bm
d}\cdot{\bm g}_{{\bm k}\lambda}({\bm r}_{1})\Bigr)\Bigl({\bm
d}\cdot{\bm g}_{{\bm k}\lambda}({\bm
r}_{2})\Bigr)\delta(\omega_{ba}-\omega).
\end{align}
\\
The parameter $\delta^{(12)}$ represents coherent coupling through
the vacuum, expressed by the frequency shift
\\
\begin{align}
\delta^{(12)}=-\frac{c}{4\pi}\mathcal{P}\int_{-\infty}^{+\infty}\frac{\omega^3}{\omega_{ba}^3}\Gamma^{(12)}(\frac{1}{\omega-\omega_{ba}}+\frac{1}{\omega+\omega_{ba}})d\omega,
\end{align}
\\
where $\mathcal{P}$ stands for Cauchy's Principal Value of the
integral. Using standard evaluating techniques and replacing
\\
\begin{align}
\sum_{\mathbf{k}\lambda}\rightarrow\sum_{\lambda}\frac{L^3}{8\pi^3}\int_{0}^{+\infty}k^2dk\int_{0}^{2\pi}d\phi\int_{0}^{\pi}\sin(\theta)d\theta,
\end{align}
\\
we find, in accordance \footnote{note a small printing error in
\cite{U Akram} in the expression for $\Gamma^{(12)}$} with \cite{U
Akram}
\\
\begin{align}
\delta^{(12)}&=\frac{3}{4}\Gamma_{ba}\Bigl(-\frac{1}{x}\cos(x)+\frac{1}{x^2}\sin(x)+\frac{1}{x^3}\cos(x)\Bigr),\nonumber\\\nonumber\\
\Gamma^{(12)}&=\frac{3}{2}\Gamma_{ba}\Bigl(\frac{1}{x}\sin(x)+\frac{1}{x^2}\cos(x)-\frac{1}{x^3}\sin(x)\Bigr)\label{delta_and_Gamma},
\end{align}
\\
where we defined the dimensionless parameter $x\equiv
\omega_{ba}r_{12}/c$, and where we used the fact that ${\bm
\mu}\bot{\bm r}_{12}$. Both couplings $\delta^{(12)}$ and
$\Gamma^{(12)}$ depend strongly on the distance between the atoms
and the orientation of $\bm \mu$ and ${\bm r}_{12}$. In Figure
\ref{fig,delta_gamma} we show how $\delta^{(12)}$ and
$\Gamma^{(12)}$ evolve as a function of $x$. For large separations
$(x\gg 1)$, the two couplings $\delta^{(12)}$ and $\Gamma^{(12)}$
vanish, independent of the orientation of the dipole moments.
However, for small separations $(x\ll 1)$, $\Gamma^{(12)}$ and
$\delta^{(12)}$ exhibit a different behavior: while
$\delta^{(12)}$ diverges for small separations, $\Gamma^{(12)}$
reduces to the single-atom decay rate $\Gamma$.
\begin{figure}
    \subfigure[]{
    \label{fig,delta}
    \includegraphics[width=3in]{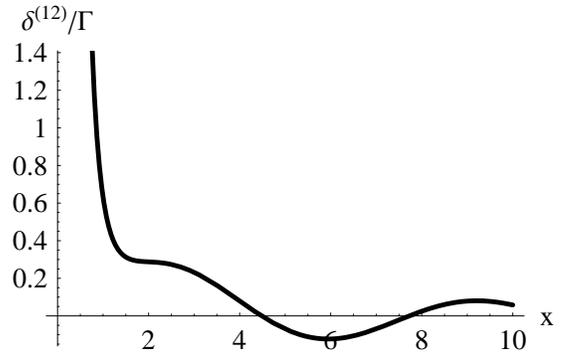}
    }
    \subfigure[]{
    \label{fig,gamma}
    \includegraphics[width=3in]{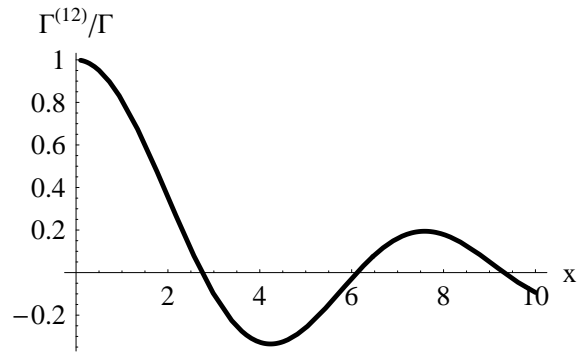}
    }
    \caption{\label{fig,delta_gamma}The coherent coupling (Figure a) and the
    incoherent coupling (Figure b) (in units of $\Gamma$) as a function of the dimensionless
    parameter $x\equiv \omega_{ba}r_{12}/c$. We have chosen ${\bm \mu}$
    perpendicular to ${\bm r}_{12}$, which is the case in the
    system studied in this paper.
    }
\end{figure}
\\
\indent The standard method of solving the Master equation
(\ref{Master,Lindblad}) is to calculate equations of motion for
the density matrix elements and solve them by direct integration
\cite{L Mandel, R Lehmberg 2}. This procedure means a set of 16
linear differential equations has to be solved. Using the
resulting density matrix, we will calculate the total field,
resulting from the scattering of ${\bm E}_{0}$. We therefore need
to establish a relation between the total field and the calculated
density matrix $\hat{\sigma}(t)$. Before establishing such
connection, we note that from a didactic point of view, it is
easier to connect both methods described in this paper if we
calculate the total field in the far-field zone, i.e., $|{\bm
r}|\gg r_{12},(\omega_{ba}/c)^{-1}$. We stress, however, that our
conclusions are also valid for observation points ${\bm r}$ in the
near-field zone. The total electric field operator is given in the
far-field by \cite{M Scully}
\\
\begin{align}
\hat{\mathbf{E}}(\mathbf{r},t)&\equiv
\hat{\mathbf{E}}^{+}(\mathbf{r},t)+\hat{\mathbf{E}}^{-}(\mathbf{r},t)\nonumber\\
&=\Biggl[\frac{1}{4\pi\varepsilon_{0}}\Bigl(\frac{\omega_{ba}}{c}\Bigr)^2
\sum_{i=1,2}\frac{1}{|\mathbf{r}-\mathbf{r}_{i}|}(\mathbf{d}_{ab}-\mathbf{R}_{i}\frac{(\mathbf{R}_{i}\cdot\mathbf{d}_{ab})}{R_{i}^2})\times\nonumber\\
&\qquad\qquad\qquad\qquad\hat{S}_{i}^{-}(t-|\mathbf{r}-\mathbf{r}_{i}|/c)\Biggr]+\text{H.c.},
\end{align}
\\
where the shorthand notation $\mathbf{R}_{i}\equiv
\mathbf{r}-\mathbf{r}_{i}$ has been used. Taking the expectation
value and Fourier transforming yields
\\
\begin{align}
\left<\hat{\mathbf{E}}[\mathbf{r},\omega]\right>&=\int_{-\infty}^{+\infty}dt\left<\hat{\mathbf{E}}(\mathbf{r},t)\right>e^{i\omega
t}\nonumber\\
&=\Bigl(\frac{\omega_{ba}}{c}\Bigr)^2
\sum_{i=1,2}\frac{1}{|\mathbf{r}-\mathbf{r}_{i}|}(\mathbf{d}_{ab}-\mathbf{R}_{i}\frac{(\mathbf{R}_{i}\cdot\mathbf{d}_{ab})}{R_{i}^2})\times\nonumber\\
&\qquad \frac{e^{i\frac{\omega}{c}|\mathbf{r}-\mathbf{r}_{i}|
}}{4\pi\varepsilon_{0}}\int_{-\infty}^{+\infty}dt e^{i\omega
t}\Bigl(\left<\hat{S}_{i}^{-}(t)\right>+\text{H.c.}\Bigr).
\end{align}
\\
If we now write (using the Green's funtion
$\overleftrightarrow{G}_{0}$ defined as
(\ref{Tmatrix,Green_Dyadic}))
\\
\begin{align}
(\mathbf{d}_{ab}-\mathbf{R}_{i}\frac{(\mathbf{R}_{i}\cdot\mathbf{d}_{ab})}{R_{i}^2})&=-4\pi
|\mathbf{r}|d_{ab}e^{-i\frac{\omega}{c}|\mathbf{r}|}\overleftrightarrow{G}_{0}(\mathbf{r}-\mathbf{r}_{i},\omega)\cdot{\bm
\mu},
\end{align}
\\
then
\\
\begin{align}
\left<\hat{\mathbf{E}}[\mathbf{r},\omega]\right>&=-\frac{d_{ab}}{\varepsilon_{0}}\Bigl(\frac{\omega_{ba}}{c}\Bigr)^2
\sum_{i=1,2}(\overleftrightarrow{G}_{0}(\mathbf{r}-\mathbf{r}_{i},\omega)\cdot{\bm
\mu})\times\nonumber\\
&\qquad\int_{-\infty}^{+\infty}dte^{i\omega
t}\Bigl(\left<\hat{S}_{i}^{-}(t)\right>+\text{H.c.}\Bigr)\label{TotalE,3}.
\end{align}
\\
In the steady-state regime, the coherences in the system (the
off-diagonal elements of the density matrix) oscillate in phase
with the incident field. We can easily study the behavior of the
coherences by transforming to the interaction picture
\\
\begin{align}
\hat{\mathcal{S}}_{i}^{+}&\equiv
e^{-i\omega_{0}t}\hat{S}_{i}^{+},\nonumber\\
\hat{\mathcal{S}}_{i}^{-}&\equiv
e^{+i\omega_{0}t}\hat{S}_{i}^{-},\qquad i\in\{1,2\},
\end{align}
\\
such that the expectation values
$\left<\hat{\mathcal{S}}_{i}^{\pm}\right>$ are time-independent in
the steady-state regime (see, e.g., \cite{C Cohen-Tannoudji}). The
field component at $\omega=\omega_{0}$ (the frequency of the
incident field) is then given by
\\
\begin{align}
\mathbf{E}[\mathbf{r},\omega_{0}]&\equiv\left<\hat{\mathbf{E}}[\mathbf{r},\omega_{0}]\right>\nonumber\\
&=-\frac{d_{ab}}{\varepsilon_{0}}\Bigl(\frac{\omega_{ba}}{c}\Bigr)^2
\sum_{i=1,2}(\overleftrightarrow{G}_{0}(\mathbf{r}-\mathbf{r}_{i},\omega)\cdot{\bm
\mu})\left<\hat{\mathcal{S}}_{i}^{-}\right>\label{TotalE,final,density}.
\end{align}
\\
Expression (\ref{TotalE,final,density}) expresses the total field
${\bm E}$ on resonance as a function of the scattering properties
of the atoms, without making any assumptions regarding the
independence of both scatterers.
\section{The T-matrix approach to scattering of light}
We now reconsider the system shown in Figure \ref{fig,setup}. We
model both atoms as point-dipoles with dipole moments given by
(\ref{dipole moments}), and study the dynamics of the system using
a T-matrix approach instead of a density matrix approach. The
total field $ {\bm E} $ at any position in the three-dimensional
coordinate space can be written as the sum of the incident field
and the field scattered by both atoms \cite{P d Vries}:
\\
\begin{align}
{\bm E}[\omega,{\bm r}]& ={\bm E}_{0}[\omega,{\bm r}]\nonumber\\
&\quad + \int \int \tensor{G}_{0}(\omega,{\bm r}-{\bm
s})\tensor{T}(\omega,{\bm s},{\bm u}){\bm E}_{0}[\omega,{\bm
u}]{\bm d}{\bm s}{\bm d}{\bm u},\label{totalE}
\end{align}
\\
where the integrations are over the full three-dimensional
coordinate space. The $3\times3$ dyadic Green's function
$\tensor{G}_{0}(\omega,{\bm r})$ has as representation in
coordinate space \cite{C Tai}:
\\
  \begin{align}
    &\tensor{G}_{0}(\omega,{\bm r}) \nonumber\\
    &\quad = -(I^{(3)}+\frac{1}{\omega^2/c^2}{\bm \nabla}\otimes{\bm \nabla})\frac{e^{i \frac{\omega}{c} r}}{4\pi
    r}\nonumber\\
      & \quad = -\frac{e^{i \frac{\omega}{c} r}}{4\pi
      r}\Bigl(P(i\frac{\omega}{c}r)I^{(3)}+\tilde{P}(i\frac{\omega}{c}r){\bm r}\otimes{\bm r}/r^2\Bigr),\label{Tmatrix,Green_Dyadic}
  \end{align}
with
\\
  \begin{align}
    P(z) = 1-\frac{1}{z}+\frac{1}{z^2}\qquad \text{and} \qquad\tilde{P}(z) =
    -1+\frac{3}{z}-\frac{3}{z^2},\label{P and Pprime}
  \end{align}
\\
where $\otimes$ denotes the tensor product. The $3\times3$ unit
tensor is denoted as $I^{(3)}$. We note that the Green's
(\ref{Tmatrix,Green_Dyadic}) function is related through (\ref{P
and Pprime}) to the coherent coupling $\delta^{(12)}$ and
incoherent coupling $\Gamma^{(12)}$ which arise in a master
equation approach (see expressions (\ref{delta_and_Gamma})) by
\\
\begin{align}
{\bm \mu}\cdot\tensor{G}_{0}(\omega,{\bm r}_{1}-{\bm
r}_{2})\cdot{\bm \mu}\equiv
\frac{1}{6\pi}\frac{\omega_{ba}/c}{\Gamma_{ba}}
\Bigl(2\delta^{(12)}-i\Gamma^{(12)}\Bigr),
\end{align}
\\
indicating that coupling through
the vacuum is intimately connected to ordinary free-space
propagation. The T-matrix $\tensor{T}$ of
the total scattering system is given by
\\
\begin{align}
\tensor{T}(\omega,{\bm r},{\bm r}')&\equiv\bra{{\bm
r}}\tensor{T}(\omega)\ket{{\bm r}'}\nonumber\\
& = {\bm \mu}\otimes{\bm \mu} \thickspace t(\omega)\sum_{i,j=1,2}
\delta
    ({\bm r}-{\bm r}_{i}) \delta({\bm r}'-{\bm r}_{j})\times\nonumber\\
&\qquad([ I^{(2)}- \tensor{D}(\omega,{\bm r}_{1},{\bm r}_{2})
t(\omega)]^{-1})_{ij},\label{totalT}
  \end{align}
\\
where the delta-functions clearly depict the local character of
both scatterers. The the $2\times2$ unit tensor is denoted by
$I^{(2)}$. The $2\times2$ matrix $\tensor{D}$ is given by
\\
\begin{align}
    &    \tensor{D}(\omega,{\bm r}_{1},{\bm r}_{2})\equiv {\bm \mu}\cdot\tensor{G}_{0}(\omega,{\bm r}_{1},{\bm r}_{2})\cdot{\bm \mu}\begin{pmatrix}
        0&1\\
        1&0\\
      \end{pmatrix}.
  \end{align}
\\
Each atom has a T-matrix element $t(\omega)$ describing its
scattering properties. In the system we are considering the
T-matrix elements of both atoms are equal because the atoms are
identical. In the limit of low-intensity incident fields, the
T-matrix element $t(\omega)$ is given by the well-known expression
for the dynamic polarizability of a two-level atom \cite{L Allen}
\\
\begin{align}
t(\omega) &=
\frac{3\pi}{\omega_{ba}/c}\frac{\Gamma_{ba}}{\Bigl(\omega-\omega_{ba}+i\frac{\Gamma_{ba}}{2}\Bigr)}\label{T,linear}.
\end{align}
\\
For incident waves of higher intensities, both atoms will show
saturation effects. The extension of expression (\ref{T,linear})
beyond the low-intensity limit is given by \cite{T Savels, P
Zoller}
\\
\begin{align}
t(\omega) &=
\frac{3\pi}{\omega_{ba}/c}\frac{\Gamma_{ba}}{\Bigl(\omega-\omega_{ba}+i\frac{\Gamma_{ba}}{2}\Bigr)}\frac{1}{1+s}\label{T,nonlinear},
\end{align}
\\
which takes into account the loss of photons at frequency $\omega$
due to inelastic scattering; the inelastically scattered photons
at frequencies $\omega'\neq \omega$ are not described. The
saturation parameter $s$ is defined as (see, e.g., \cite{C
Cohen-Tannoudji})
\\
\begin{align}
s\equiv\frac{\Omega^2/2}{(\omega_{ba}-\omega)^2+\Gamma^2/4},
\end{align}
\\
with the Rabi frequency $\Omega\equiv |{\bm d}_{ab}\cdot {\bm
E}_{0}|/\hbar$. With the T-matrix element (\ref{T,linear}) we can
rewrite the total field (\ref{totalE}) as
\\
\begin{align}
{\bm E}[\omega,{\bm r}]=&{\bm E}_{0}[\omega,{\bm r}]+
\frac{t(\omega)}{1-t(\omega){\bm
\mu}\cdot\tensor{G}_{0}(\omega,{\bm r}_{1},{\bm
r}_{2})\cdot{\bm \mu}}\times\nonumber\\
&\sum_{i,j=1,2}\tensor{G}_{0}(\omega,{\bm r}-{\bm r}_{i})\cdot
    {\bm \mu}\otimes{\bm
    \mu}\cdot{\bm E}_{0}[\omega,{\bm r}_{j}]\label{TotalE,2}.
\end{align}
\\
Equation (\ref{TotalE,2}) is the result of a T-matrix calculation,
expressing the total field ${\bm E}$ as a function of the incident
field ${\bm E}_{0}$ and the scattering properties of both atoms,
explicitly assuming that they are independent from one another.
\section{Comparison of both methods and Discussion}
We now compare expressions (\ref{TotalE,final,density}) and
(\ref{TotalE,2}) and discuss their discrepancies. We consider
three regimes.
\subsection{Low-saturation incident field}
If the saturation is very small ($s \ll 1$), the steady-state
value of the density matrix can be calculated analytically. The
resulting steady-state expectation value
$\left<\hat{\mathcal{S}}_{i}^{-}\right>$, $i\in \{1,2\}$ is, using
the linear expression (\ref{T,linear}),
\\
\begin{align}
\lim_{\Omega\rightarrow
0}\frac{1}{\Omega}\left<\hat{\mathcal{S}}_{i}^{-}\right>
&=\frac{\omega_{ba}/c}{6\pi\Gamma_{ba}}\frac{t(\omega)}{1-t(\omega){\bm
\mu}\cdot\tensor{G}_{0}(\omega,{\bm r}_{1},{\bm r}_{2})\cdot{\bm
\mu}},
\end{align}
\\
which yields, if substituted in (\ref{TotalE,final,density}), the
same expression (\ref{TotalE,2}) as given by a T-matrix approach,
hereby proving the validity of the T-matrix formalism for
low-intensity incident fields for arbitrary separation between
both atoms.
\subsection{Atoms in each other's far-field: $r_{12}\gg (\omega_{ba}/c)^{-1}$}
In the limit of large separation between both atoms
$r_{12}\gg(\omega_{ba}/c)^{-1}$, both the coherent coupling
$\delta^{(12)}$ and the incoherent coupling $\Gamma^{(12)}$ reduce
to zero. In this regime, the density matrix $\hat{\sigma}$ obeys
\\
\begin{align}
\text{Tr}_{1}(\hat{\sigma})=\text{Tr}_{2}(\hat{\sigma})=\sigma_{0},\label{Tr1=Tr2}
\end{align}
\\
where $\sigma_{0}$ is the ($2\times 2$) density matrix
corresponding to a single two-level system (see, e.g., \cite{C
Cohen-Tannoudji}), and $\text{Tr}_{i}$ stands for the trace over
atom $i$. In other words, if both atoms are in each other's
far-field, the single-atom density matrix factorizes out in two
single-atom density matrices. This result corresponds with the
intuitive expectation that the independent scattering
approximation (\ref{TotalE,2})  is valid for large separation.
\begin{figure}[t]
    \subfigure[$r_{12}=100 (\omega_{ba}/c)^{-1}$]{
    \label{E_far,Re,delta}
    \includegraphics[width=3in]{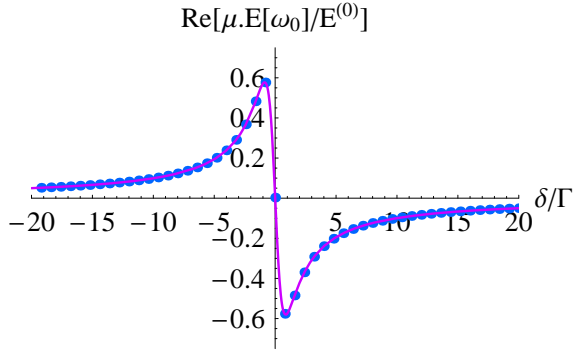}
    }
    \subfigure[$r_{12}=0.8(\omega_{ba}/c)^{-1}$]{
    \label{E_close,Re,delta}
    \includegraphics[width=3in]{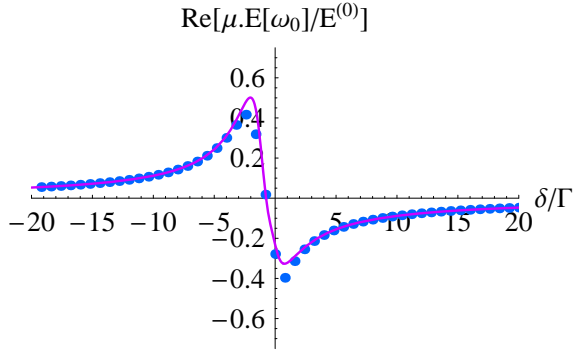}
    }
    \caption{\label{E,Re,delta}The real part of the field component
    ${\bm \mu}\cdot{\bm E}[\omega_{0}]/E^{(0)}$ on resonance,
    evaluated at the perpendicular bisector of ${\bm r}_{12}$.
    The Rabi frequency of the incident field is chosen $\Omega/\Gamma=1$.
    The field has been calculated using both the T-matrix formalism (solid line)
    and the density matrix formalism (dotted line).
    The field has been plotted as a function of the detuning $\delta$,
    for (a) atoms in each other's far-field, and for (b) atoms
    in each other's near-field.}
\end{figure}
\begin{figure}[t]
    \subfigure[$r_{12}=100 (\omega_{ba}/c)^{-1}$]{
    \label{E_far,Im,delta}
    \includegraphics[width=3in]{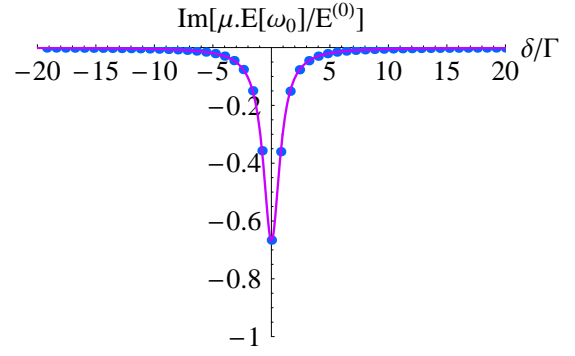}
    }
    \subfigure[$r_{12}=0.8(\omega_{ba}/c)^{-1}$]{
    \label{E_close,Im,delta}
    \includegraphics[width=3in]{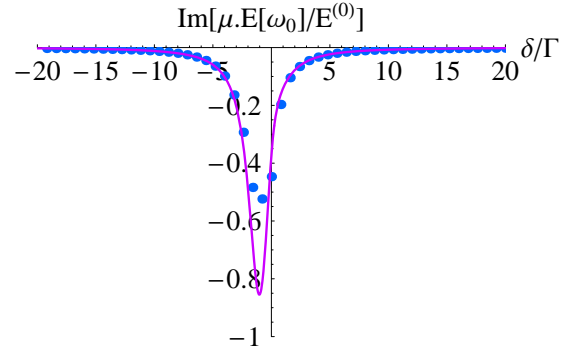}
    }
    \caption{\label{E,Im,delta}The imaginary part of the field component
    ${\bm \mu}\cdot{\bm E}[\omega_{0}]/E^{(0)}$ on resonance,
    evaluated at the perpendicular bisector of ${\bm r}_{12}$.
    The Rabi frequency of the incident field is chosen $\Omega/\Gamma=1$.
    The field has been calculated using both the T-matrix formalism (solid line)
    and the density matrix formalism (dotted line).
    The field has been plotted as a function of the detuning $\delta$,
    for (a) atoms in each other's far-field, and for (b) atoms
    in each other's near-field.}
\end{figure}
\subsection{Arbitrary separation and incident field in the saturated regime}
If the saturation parameter $s$ is of order $1$ or larger and the
separation between the atoms is of the order of (or smaller than)
$(\omega_{ba}/c)^{-1}$, no easy analytic expressions for the
density matrix exist. However, the density matrix and the total
field can still be evaluated numerically. In Figure
\ref{E,Re,delta} - \ref{E,Im,omega}, we have calculated ${\bm
\mu}\cdot{\bm E}[\omega_{0}, {\bm r}]$ using expressions
(\ref{TotalE,final,density}) and (\ref{TotalE,2}). Obviously, the
numerical value of the total field component ${\bm \mu}\cdot{\bm
E}[\omega_{0}, {\bm r}]$ depends on numerous parameters such as
the incident field strength, the distance between both atoms, the
evaluation point ${\bm r}$ and the frequency of the incident
field. It is not our intention, however, to fully explore the
entire $(\omega_{ba}-\omega_{0},{\bm r}, r_{12}, \Omega)$
parameter space. We only want to point out that differences
between both
methods do arise, and explain why.\\
\indent As an example, the field has been evaluated at the
perpendicular bisector of ${\bm r}_{12}$. The field is shown in
units of
\\
\begin{align}
{\bm E}^{(0)}\equiv\frac{6\pi}{\omega_{ba}/c}{\bm
\mu}\cdot\Bigl(\tensor{G}_{0}({\bm r}-{\bm
r}_{1})+\tensor{G}_{0}({\bm r}-{\bm r}_{2})\Bigr)\cdot{\bm \mu}.
\end{align}
\\
Figure \ref{E,Re,delta} and \ref{E,Im,delta} show the field as a
function of the detuning $\delta\equiv 2(\omega_{ba}-\omega_{0})$.
Figure \ref{E,Re,omega} and \ref{E,Im,omega} show the field as a
function of the Rabi frequency $\Omega$, expressing the intensity
of the incident field. It is clear that if both atoms are in each
other's far-field ($r_{12}\gg(\omega_{ba}/c)^{-1}$), or if the
saturation parameter is very small ($s\ll 1$), the predictions of
the T-matrix approach and the density matrix approach coincide, as
explained above. However, we see that for atoms in each other's
near-field ($r_{12}\approx(\omega_{ba}/c)^{-1}$), the predictions
of both approaches deviate for increasing saturation parameter of
the incident field. At very high saturation ($s\gg
1$), both approaches again yield the same results.\\
\indent We can explain the difference between both approaches as
follows. For low-intensity incident fields, both atoms scatter
light purely elastically. At higher intensities, more light is
scattered inelastically (and less light elastically). In a
T-matrix approach, the coupling of both atoms with inelastically
scattered light is not taken into account. This explains why, in
the case of a single scatterer, the T-matrix approach is exact,
why for multiple scatterers, problems can arise when using
T-matrices. The reason for this is that in the regime where the
T-matrix approach breaks down, it is impossible to quantify the
incident light on each scatterer, since the incident light to a
certain extend consists of incoherent light, generated by multiple
inelastic scattering between both atoms. It is then no longer
possible to define a single-scatterer scattering property, since
coherences arise between different scatterers, which are induced
by inelastically scattered light. Finally, at very high
intensities of the incident field, both atoms will be completely
saturated, regardless of the approach one uses to calculate the
field. This saturation explains why for large $s$, the results
produced by both methods again coincide.
\\
\begin{figure}[t]
    \subfigure[$r_{12}=100 (\omega_{ba}/c)^{-1}$]{
    \label{E_far,Re,omega}
    \includegraphics[width=3in]{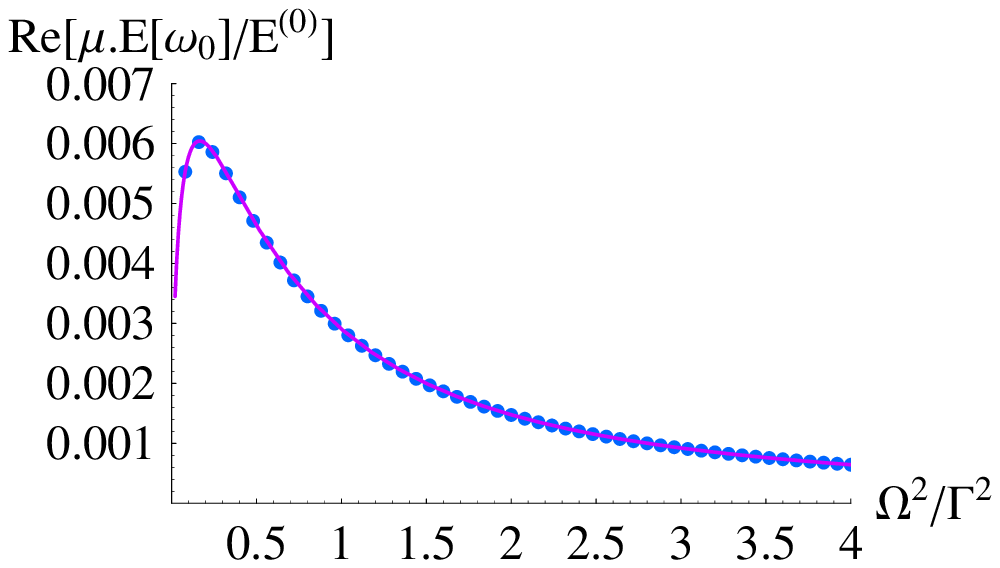}
    }
    \subfigure[$r_{12}=0.8(\omega_{ba}/c)^{-1}$]{
    \label{E_close,Re,omega}
    \includegraphics[width=3in]{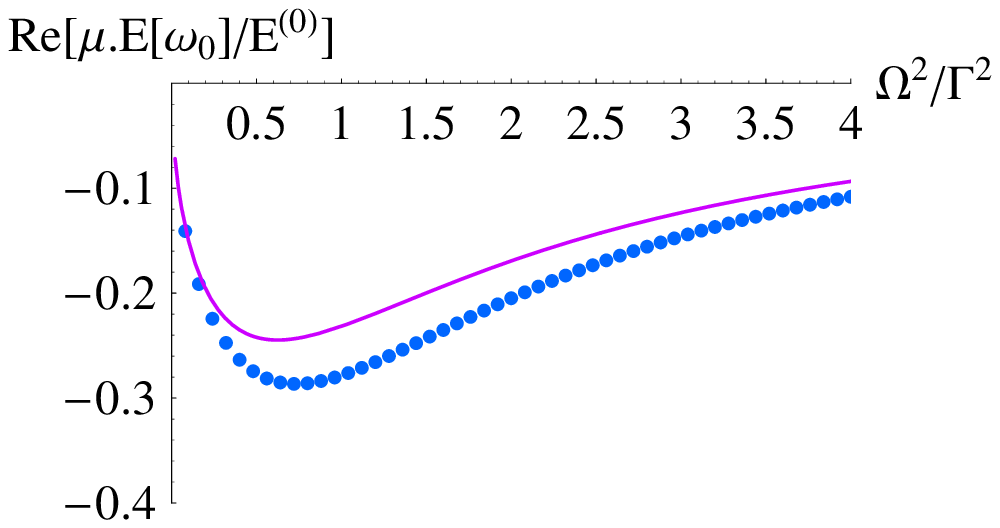}
    }
    \caption{\label{E,Re,omega}The real part of the field component
    ${\bm \mu}\cdot{\bm E}[\omega_{0}]/E^{(0)}$ on resonance,
    evaluated at the perpendicular bisector of ${\bm r}_{12}$.
    The detuning of the incident field is chosen $\delta=0$.
    The field has been calculated using both the T-matrix formalism (solid line)
    and the density matrix formalism (dotted line).
    The field has been plotted as a function of the Rabi frequency $\Omega$,
    for (a) atoms in each other's far-field, and for (b) atoms
    in each other's near-field.}
\end{figure}
\begin{figure}[t]
    \subfigure[$r_{12}=100 (\omega_{ba}/c)^{-1}$]{
    \label{E_far,Im,omega}
    \includegraphics[width=3in]{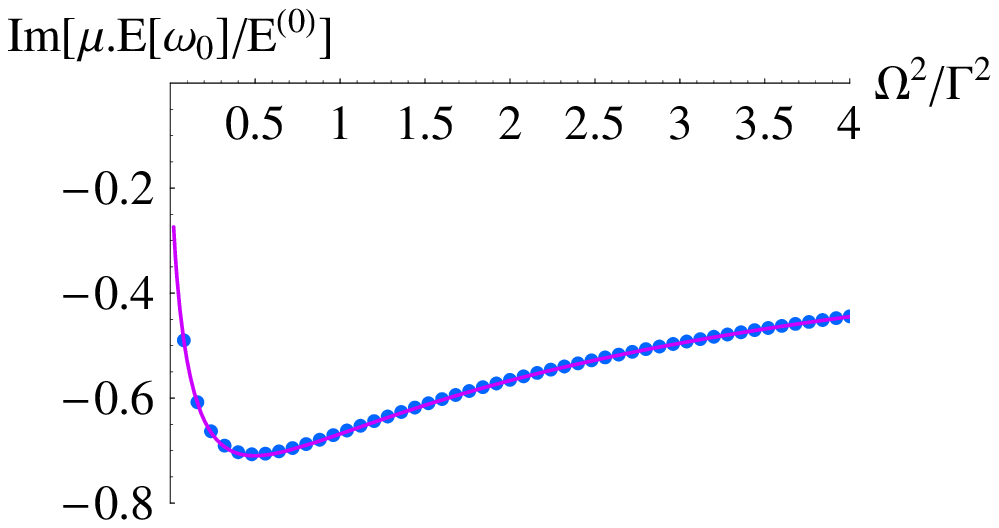}
    }
    \subfigure[$r_{12}=0.8(\omega_{ba}/c)^{-1}$]{
    \label{E_close,Im,omega}
    \includegraphics[width=3in]{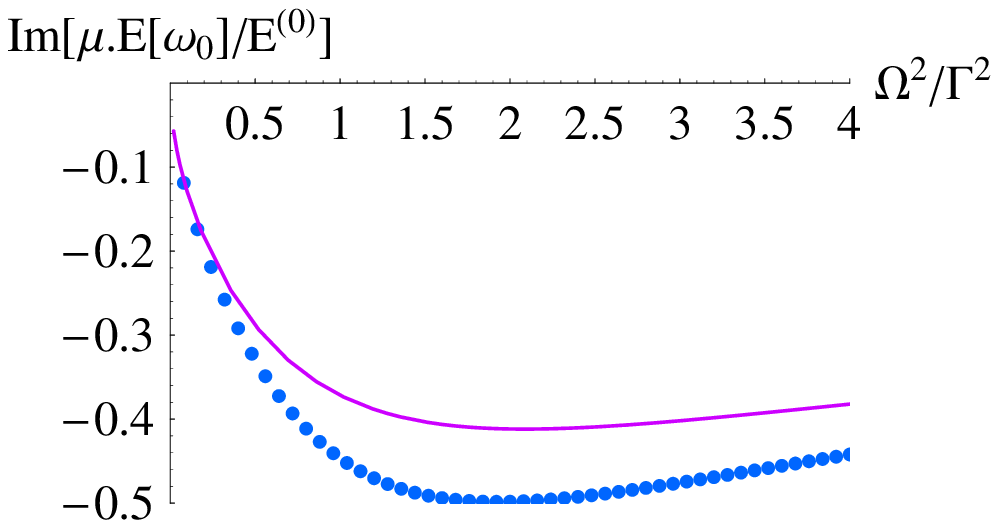}
    }
    \caption{\label{E,Im,omega}The imaginary part of the field component
    ${\bm \mu}\cdot{\bm E}[\omega_{0}]/E^{(0)}$ on resonance,
    evaluated at the perpendicular bisector of ${\bm r}_{12}$.
    The detuning of the incident field is chosen $\delta=0$.
    The field has been calculated using both the T-matrix formalism (solid line)
    and the density matrix formalism (dotted line).
    The field has been plotted as a function of the Rabi frequency $\Omega$,
    for (a) atoms in each other's far-field, and for (b) atoms
    in each other's near-field.}
\end{figure}
\section{Extension}
\indent We stress that the breakdown of the independent scattering
approximation we describe here applies to many other systems. In
general, all systems where the contribution of incoherent light is
not negligible will experience a breakdown of the independent
scattering approximation. If, for example, one wishes to describe
atomic systems with gain \cite{T Savels} or the effect of
saturation on coherent backscattering \cite{A Buchleitner, T
Wellens, T Wellens 2} using independent atoms, one is limited to
large separations between the atoms. If one wishes to describe
these systems for atoms in each other's near-field, then the atoms
cannot be described (from a light scattering point of view) as
individual entities; they are to be considered as one large
indivisible system. Only a description taking into account all
self-induced internal coherences, such as
a density-matrix description, then produces correct results.\\
\section{Summary}
In this paper, we studied the scattering properties of a system
consisting of two atoms in free space. The scattering of incident
monochromatic light has been considered, using both a saturable
single-frequency elastic T-matrix approach and a density matrix
approach. The former method assumes both atoms to be independent,
while the latter method does not make such initial assumption. We
found that in the regime of low saturation parameters, or for
large distance between both atoms, both approaches yield the same
predictions. However, if the atoms are in each other's near-field
and the saturation parameter is high enough, the two approaches
produce different results. This discrepancy is due inelastic
scattering of light, which is not taken into account in a T-matrix
approach.
\begin{acknowledgements}
This work is part of the research program of the `Stichting voor
Fundamenteel Onderzoek der Materie' (FOM), which is financially
supported by the `Nederlandse Organisatie voor
Wetenschappelijk Onderzoek' (NWO).\\
\end{acknowledgements}


\begin{thebibliography}{99'}
\bibitem{B v Tiggelen}
A. Lagendijk and B. A. van Tiggelen, Physics Rep. \textbf{270},
143 (1996)
\bibitem{Th Nieuwenhuizen}
Th. M. Nieuwenhuizen, A. Lagendijk and B.A. van Tiggelen, Phys.
Lett. A \textbf{169}, 191 (1992)
\bibitem{C Bohren}
C.F. Bohren and D.R. Huffman, \textit{Absorption and scattering of light by small particles} (Wiley Science Paper Series, New
York, 1998)
\bibitem{M Havey}
D.V. Kupriyanov, I.M. Sokolov, P. Kulatunga, C.I. Sukenik and M.D.
Havey, Phys. Rev. A \textbf{67}, 013814  (2003)
\bibitem{M Mark}
M. B. van der Mark, M. P. van Albada and Ad Lagendijk, Phys. Rev.
B, \textbf{37}, 3575 (1988)
\bibitem{T Chaneliere}
T. Chaneli\`{e}re, D. Wilkowski, Y. Bidel, R. Kaiser and C.
Miniatura, Phys. Rev. E \textbf{70}, 036602 (2004)
\bibitem{D Gammon}
D. Gammon and D.G. Steel, Phys. Today \textbf{55}, 36 (2002)
\bibitem{D Leibfried}
D. Leibfried, R. Blatt, C. Monroe and D. Wineland, Rev. Mod. Phys. \textbf{75}, 281 (2003)
\bibitem{C Cohen-Tannoudji}
C. Cohen-Tannoudji, J. Dupont-Roc and G. Grynberg,
\textit{Atom-Photon Interactions} (Wiley Science Paper Series, New
York, 1998)
\bibitem{M Scully}
M. O. Scully and M. S. Zubairy, \textit{Quantum Optics} (Cambridge
University Press, Cambridge, 1997)
\bibitem{H Carmichael}
H.J. Carmichael, \textit{Statistical Methods in Quantum Optics 1}
(Springer-Verlag Berlin Heidelberg New York, 2002)
\bibitem{R Lehmberg}
R. H. Lehmberg, Phys. Rev. A \textbf{2}, 883 (1970)
\bibitem{G Lindblad}
G. Lindblad, Commun. Math. Phys. \textbf{48}, 119 (1976)
\bibitem{U Akram}
U. Akram, Z. Ficek and S. Swain, Phys. Rev. A \textbf{62}, 013413
(2000)
\bibitem{D Cardimona}
D.A. Cardimona and C.R. Stround, Jr., Phys. Rev. A \textbf{27}, 2456 (1983)
\bibitem{P Milonni}
P. W. Milonni and P. L. Knight, Phys. Rev. A \textbf{10}, 1096 (1974)
\bibitem{P Milonni 2}
P. W. Milonni and P. L. Knight, Phys. Rev. A \textbf{11}, 1090
(1975)
\bibitem{L Mandel}
L. Mandel and E. Wolf, \textit{Optical Coherence and Quantum Optics} (Cambridge Univ. Press, Cambridge, 1995)
\bibitem{R Lehmberg 2}
R. H. Lehmberg, Phys. Rev. A \textbf{2}, 889 (1970)
\bibitem{P d Vries}
P. de Vries, D. V. van Coevorden and A. Lagendijk, Point
scatterers for classical waves, Rev. Mod. Phys. \textbf{70}, 1980
\bibitem{C Tai}
C.T. Tai, \textit{Dyadic Green Functions in Electromagnetic
Theory} (IEEE Press, New York, 1994)
\bibitem{L Allen}
L. Allen and J. H. Eberly, \textit{Optical Resonance and Two-level
Atoms} (Dover Publications, New York, 1987)
\bibitem{T Savels} T. Savels, A.P. Mosk and Ad Lagendijk,
\textit{Light scattering from three-level systems: The T-matrix of
a point-dipole with gain}, Phys. Rev. A, in press (2005)
\bibitem{P Zoller}
P. Zoller, Phys. Rev. A, \textbf{20}, 2420 (1979)
\bibitem{A Buchleitner}
V. Shatokhin, C. A. M{\"u}ller, and A. Buchleitner, Phys. Rev.
Lett. \textbf{94}, 043603 (2005)
\bibitem{T Wellens}
T. Wellens, B. Gr{\'e}maud, D. Delande, and C. Miniatura, Phys.
Rev. A \textbf{70}, 023817 (2004)
\bibitem{T Wellens 2}
 T. Wellens, B. Gr{\'e}maud, D. Delande, C. Miniatura, cond-mat/0411555
\end{thebibliography}
\end{document}